\documentstyle[aps,prb,twocolumn,epsf]{revtex}

\begin{document}
\draft
\wideabs{
\title{Non-equilibrium tunneling into general quantum Hall edge states}
\author{Joel E. Moore$^a$, Prashant Sharma$^b$, and Claudio Chamon$^{b,c}$}
\address{$^a$ Department of Physics, Massachusetts Institute of Technology,
Cambridge MA 02139 \\
$^b$ Department of Physics, Boston University, Boston, MA 02215 \\
$^c$ School of Natural Sciences, Institute for Advanced Study,
Princeton NJ 08540}
\maketitle
\begin{abstract}
  In this paper we formulate the theory of tunneling into general Abelian
  fractional quantum Hall edge states. In contrast to the simple Laughlin
  states, a number of charge transfer processes must be accounted for.
  Nonetheless, it is possible to identify a unique value corresponding to
  dissipationless transport as the asymptotic large-$V$ conductance through a
  tunneling junction, and find fixed points (CFT boundary conditions)
  corresponding to this value.  The symmetries of a given edge tunneling
  problem determine the appropriate boundary condition, and the boundary
  condition determines the strong-coupling operator content and current
  noise.
\end{abstract}
}

\section{Introduction}
\label{sec:intro}

Tunneling into the edges of fractional quantum Hall (FQH) states is important
as the most experimentally accessible probe of the fractionally charged
quasiparticles believed to exist in bulk FQH states.  Although a large body
of work has been carried out on tunneling between edges of Laughlin states,
and more generally on tunneling between Luttinger liquids, a theoretical
understanding of tunneling into general FQH states is still missing beyond
weak coupling, where the comparison to experiment is puzzling.  As opposed to
the edges of Laughlin states ($\nu=\frac{1}{2q+1}$), general Abelian edge
states such as the main sequence $\nu=\frac{p}{2pq+1}$ contain several
quasiparticle and electron operators and thus multiple tunneling processes
transferring different amounts of charge.  The purpose of this paper is to
develop a framework to study this multiple tunneling problem.

The chiral-Luttinger-liquid ($\chi$LL) model~\cite{wen} for incompressible
filling fractions~\cite{kfp,moore} and a composite-fermion theory for
compressible states~\cite{shytov} both predict that the tunneling exponent
$\alpha$ in $I \propto V^\alpha$ for tunneling into FQH edges has a plateau
structure as a function of $\nu$.  Experiments by Grayson {\it
  et.~al.}~\cite{grayson} show a smooth dependence $\alpha \approx \nu^{-1}$
for most samples, although some samples do show a plateau structure near $\nu
= 1/3$~\cite{chang}. Different theoretical scenarios have
emerged~\cite{pasquier}, some predicting $\alpha = \nu^{-1}$. The natural
question to ask is whether these theories will also endure other experimental
tests. More precisely, these theories share the same weak-coupling
predictions, but will not necessarily agree at strong coupling. In this paper
we describe the strong coupling physics for the $\chi$LL theory of edge
states and show that the resulting value of large-$V$ conductance corresponds
to a dissipationless zero-temperature fixed point.  The result should be
comparable to experiments on point tunneling between quantum Hall states and
to other candidate theories.

The single-edge tunneling problem is known to contain a great deal of
interesting physics.  In tunneling between two $\nu=1/3$ edges, there are two
starting points: electron ($e$) or quasiparticle ($e^*=e/3$) tunneling.
These two regimes are connected by a weak-strong duality symmetry.  In each
regime, there is only a single tunneling process (or operator) that transfers
the respective charge. An exact solution~\cite{fendley} via the thermodynamic
Bethe ansatz describes completely the crossover between these two pictures,
along the integrable trajectory. In tunneling between edges of the two-mode
hierarchy state $\nu = 2/5$, there are two most relevant electron operators
(charge $e$ and fermionic statistics) in each edge, and consequently four
ways to transfer charge $e$ between the edges.  At strong coupling there are
both charge $e/5$ and charge $2e/5$ quasiparticles which can tunnel from one
edge to the other.  We will show that the important properties (conductance,
noise, operator content) near the strong-coupling fixed point can be
determined without an exact solution for the crossover.

Just as interesting as tunneling between edges of the same FQH state is the
``mismatched'' problem of tunneling between different filling fractions
$\nu_1,\nu_2$~\cite{chklovskii,chamon}, where new effective fractional
charges appear in the strongly coupled system in the single-mode
case~\cite{sandler}; this case includes tunneling from a metal (which can be
modeled by the Fermi liquid $\nu = 1$ state) to a FQH state. In this problem,
like in the case of same edge tunneling, the hierachical edge states contain
several electron operators, and therefore there are many ways of transferring
charge from the Fermi liquid reservoir to the FQH edges.

In this paper we address the problem of tunneling between hierachical edge
modes, using different approaches. We start by giving in section
\ref{sec:power} an elementary argument that two values of the junction
conductance correspond to dissipationless transport, and determine the
boundary conditions on bosonic modes in the $\chi$LL theory which correspond
to these conductances. Then a conformal field theory calculation of the
partition function is used in section \ref{sec:CFT} to justify the
strong-coupling duality picture of instantons between different minima of the
tunneling operators. This calculation determines the operator content at
strong coupling, which determines the noise and corrections to the tunneling
current. These corrections could in principle distinguish among different
candidate theories, even when they share the same asymptotic large-voltage
conductance.  Section \ref{sec:conclusions} contains a brief summary of our
results.

\section{Large-voltage conductance}
\label{sec:power}

In this section we focus on the aymptotic large-voltage conductance
for a general junction between two FQH states. We start with an
elementary argument based on energy conservation, and show that the
value suggested by this argument corresponds to specific boundary
conditions on the neutral modes in the $\chi$LL theory.  ``Neutral
modes'' here refers to those which do not carry charge in their action
on the complete system of two edges.  For example, a mode which adds
charge to one edge and removes an equal charge from the other is a
neutral mode of the whole system, even though its restriction to
either edge is charged.  In other words, neutral modes of the combined
two-edge system need not be combinations of neutral modes from each
multi-mode subedge; they may also include charged modes from each
subedge, as long as the total charge is zero.

\subsection{Dissipation and conductance}

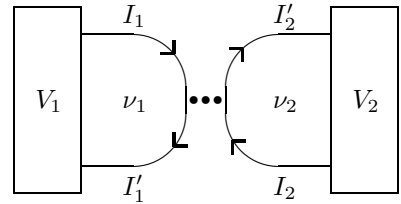
\begin{figure}
\centerline{\begin{picture}(200,95)
\put(50,50){\oval(80,50)[r]}
\put(145,50){\oval(80,50)[l]}
\put(25,15){\framebox(25,70)}
\put(145,15){\framebox(25,70)}
\put(33,47){$V_1$}
\put(153,47){$V_2$}
\put(66,47){$\nu_1$}
\put(123,47){$\nu_2$}
\put(66,79){$I_1$}
\put(66,14){$I^\prime_1$}
\put(123,79){$I^\prime_2$}
\put(123,14){$I_2$}
\put(97.5,50){\circle*{3}}
\put(93,50){\circle*{3}}
\put(102,50){\circle*{3}}
\thicklines
\put(85.5,32.5){\line(0,1){5}}
\put(85.5,32.5){\line(1,0){5}}
\put(85.5,67.5){\line(0,1){5}}
\put(85.5,67.5){\line(-1,0){5}}
\put(111.5,69.5){\line(0,-1){5}}
\put(111.5,69.5){\line(-1,0){5}}
\put(107.5,34.5){\line(0,-1){5}}
\put(107.5,34.5){\line(1,0){5}}
\thinlines
\end{picture}}
\caption{Schematic geometry for point tunneling between quantum Hall
states.}
\label{figone}
\end{figure}

A simple argument shows that an upper bound on the conductance through a
junction between FQH edges follows from the assumption that only one mode on
each edge couples to the electric potential.  In the $\chi$LL theory, this
assumption holds in the presence of either unscreened Coulomb interactions or
random hopping at the edge, for both nonchiral~\cite{kfp} and
chiral~\cite{moore2} edges. Since in experiments the Coulomb interaction is
screened only at moderate distances, and impurities are present, we assume
separation of charge and neutral modes on each edge. The currents labeled in
Fig.~\ref{figone} are $I_1 = \nu_1 V_1, I_2 = \nu_2 V_2, I_1^\prime = \nu_1
V_1^\prime, I_2^\prime = \nu_2 V_2^\prime$ (here $\frac{e^2}{h} = 1$).
Current and energy conservation at the junction give
\begin{eqnarray}
\nu_1 (V_1 - V_1^\prime) + \nu_2 (V_2 - V_2^\prime) &=& 0
\\
\nu_1 (V_1^2 - {V_1^\prime}^2) + \nu_2 (V_2^2 - {V_2^\prime}^2) &=& 2 P.
\end{eqnarray}
Here $P$ is the power dissipated at the junction.  In terms of the
two-terminal conductance $g = I_{\rm t} / (V_1 - V_2)=I_{\rm t} / V$,
\begin{equation}
P = g^2 V^2 \left[ {1 \over g} - {{\nu_1}^{-1} + {\nu_2}^{-1} \over 2}
\right].
\end{equation}

The dissipated power is zero for $g=0$ or
$g = \sigma_{hm} = 2\nu_1 \nu_2 /(\nu_1+\nu_2)$,
and positive for intermediate values; energy conservation forbids
values $g \geq \sigma_{hm}$.  The dissipated power can go into
excitations of the oscillator modes of the outgoing edges~\cite{chamon}.
The currently known fixed points for tunneling between quantum Hall
edges all have zero dissipation at zero temperature except in the presence
of exactly marginal operators (as in tunneling between $\nu = 1$ states).
We thus conjecture that, unless marginal tunneling operators are present,
the conductance saturates for large $V_1 - V_2$ at the value
$\sigma_{hm}$.

\subsection{Boundary conditions and conductance}

The two values of the dissipationless tunneling conductance result
from imposing either Neumann ($N$) or Dirichlet ($D$) boundary
conditions on one of the neutral bosonic modes in the $\chi$LL theory
of the combined (two-edge) system. The $N$ condition corresponds to no
tunneling in the geometry of Fig.~\ref{figone}, hence $g=0$, whereas
the $D$ condition saturates the upper bound $g=\sigma_{\rm hm}$.
There can be several boundary conditions with the same value of
conductance: these differ in the conformally invariant boundary
conditions of other neutral modes and in operator content. The total
charge mode always has $N$ boundary condition from charge
conservation.

An edge of a state with $n$ condensates is described by a universal $n \times
n$ matrix $K$ and charge $n$-vector ${\bf t}$ inherited from the bulk
Chern-Simons effective theory. For chiral edges (all modes
propagate in the same direction), $K$ is positive definite and the scaling
dimension of the vertex operator $O_{\bf m} = \exp(i m_j \phi_j)$ is
$\Delta({\bf m}) = \frac{1}{2}{\bf m}^{\rm T} K^{-1} {\bf m}$.  For nonchiral
edges such as $\nu = 2/3$, the same holds but with $K^{-1}$ replaced by the
scaling-dimension matrix $\Delta$~\cite{kfp,moore}.  Define an enlarged
$n_K$-dimensional $K$-matrix ($n_K = n_1 + n_2$) that combines both edges,
one with an $n_1\times n_1$ matrix $K_1$ and the other with an $n_2\times
n_2$ matrix $K_2$:
\begin{equation}
K=\left[
\matrix{
K_1 & 0\cr
0 & K_2  \cr}
\right].
\label{kdef}
\end{equation}

The conductance is obtained from the Kubo formula applied to the charge
density operator on one edge~\cite{kane,nayak}. The charge density for edge 1
is written as $\rho_{e1}=\sum_{i=1}^{n_K} t^1_i
\rho_i$, where ${\bf t}^{1}$ is the charge vector for edge 1 (notice that
${\bf t}^1$ is padded with $n_2$ zeros to length $n_K$). The total charge
mode (for the combined system) is $ \rho_{t}=\sum_{i=1}^{n_K} t_i \rho_i$
where the total charge vector ${\bf t} = {\bf t}^1 + {\bf t}^2$.  We can now
split the boson fields associated to these densities (through
$\rho=\frac{1}{2\pi}\partial_x \phi$) into charged and neutral parts:
$\phi_{e1}=\alpha \phi_{t} + \sum_{i=1}^{n_K - 1} l^{n}_i \phi_i$, or
alternatively, ${\bf t}^{1}=\alpha {\bf t} + {\bf l}^n$.
The requirement that the last term be neutral implies ~\cite{wen} that
$Q(l^n)={\bf t}^{\rm T}K^{-1}{\bf l}^n=0$, which is simply the statement that
${\bf l}^n$ corresponds to a neutral object. Then $\alpha$ is fixed since
${\bf t}^T K^{-1} {\bf t}^1=\alpha {\bf t}^T K^{-1} {\bf t} +0$.  Now,
using $\nu_1={{\bf t}^1}^T K^{-1} {\bf t}^1$ (and likewise for
$\nu_2$) and the definition of $K$ in (\ref{kdef}), we obtain
$\alpha=\nu_1/(\nu_1+\nu_2)$, and ${\bf l}^n = \nu_2 {\bf t}^1 / (\nu_1 +
\nu_2)
- \nu_1 {\bf t}^2 / (\nu_1 + \nu_2).$

The conductance of the junction is given by the difference between the
$N$ boundary condition (which corresponds to decoupled edges or zero
tunneling conductance) and the $D$ boundary condition (corresponding
to strong coupling) applied to the {\it neutral} mode ${\bf l}^n$.
This mode transfers charge from one edge to the other but conserves
total charge.  Explicitly, in terms of correlations of the boson
fields across the tunneling site ($x=0$)
\begin{equation}
g_D^1-g_N^1=2 \frac{|\omega|}{2\pi} \left(
\langle |\phi_{e_1}(\omega)|^2\rangle
{\big |}_D
-
\langle |\phi_{e_1}(\omega)|^2\rangle
{\big |}_N
\right).
\label{eq:gkubo}
\end{equation}
For the $N$ case we have:
\begin{eqnarray}
\frac{|\omega|}{2\pi}
\langle \phi_{e_1}(\omega) \phi_{e_1}(-\omega) \rangle
&=&
\alpha^2 {\bf t}^T K^{-1} {\bf t} + {{\bf l}^n}^T K^{-1} {\bf l}^n \nonumber \\
&=&
{{\bf t}^1}^T K^{-1} {\bf t}^1 = \nu_1.
\label{eq:gN}
\end{eqnarray}
Now, for the $D$ boundary condition we have
\begin{eqnarray}
\frac{|\omega|}{2\pi}
\langle \phi_{e_1}(\omega) \phi_{e_1}(-\omega) \rangle
&=&
\alpha^2 {\bf t}^T K^{-1} {\bf t} + {\rm zero} \nonumber \\
&=&
\alpha^2 (\nu_1+\nu_2)= \frac{\nu_1^2}{\nu_1+\nu_2}.
\label{eq:gD}
\end{eqnarray}
Notice that because ${\bf l}^n$ was pinned to zero ($D$ boundary
condition) its contribution to the correlation is null. It then
follows from Eqs.~(\ref{eq:gkubo},\ref{eq:gN},\ref{eq:gD}) that
the conductance through the junction in Fig.~\ref{figone}
is \begin{equation}
g = g_N^1 - g_D^1 =\frac{2\nu_1\nu_2}{\nu_1+\nu_2}=\sigma_{hm},
\end{equation}
the harmonic average of the two filling factors.  The two different boundary
conditions on ${\bf l}^n$ correspond to the two values $g=0$ and
$g=\sigma_{hm}$ for which the transport is dissipationless. The conductance
is independent of the boundary condition on neutral modes ${\bf n}$ with
${\bf n}^{\rm T} K^{-1} {\bf l}^n = 0$, and it will be shown below that in
tunneling between edges of the same state, some neutral modes retain $N$
boundary conditions at strong coupling. Because the strong coupling
conductance depends only on the neutral mode ${\bf l}^n$ formed from the
charge operators on each edge, our result may well apply in other edge
theories with the same charge mode as the $\chi$LL but different neutral
modes.  However, only in bosonic theories such as the $\chi$LL are there
known techniques (e.g., instanton expansion) to calculate properties at
strong coupling.  If other theories do become calculable at strong coupling,
there will likely be differences in operator content from the $chi$LL
predictions.

\section{Properties of the strong-coupling fixed point}
\label{sec:CFT}

We now study the strong-coupling state for two illustrative cases
(tunneling between $\nu = 1$ and $\nu = 2/5$ edges, and tunneling
between two $\nu = 2/5$ edges) using techniques which generalize
directly to other cases.  For $\nu=1$ to $\nu=2/5$ the strong-coupling
properties correspond to the above conjectured fixed point with $D$
conditions on the neutral modes of the combined system.  In a scaled
and rotated basis $(\phi_c,\phi_1,\phi_2)$ with $K = I$ and
$\phi_1,\phi_2$ neutral, the most relevant electron tunneling
operators at weak coupling are given by ${\bf m}^\pm = (0,
\sqrt{7}/2,\pm 1/2)$ with scaling dimension $\Delta({\bf m}^\pm) = 2$.
The action in this basis is
\begin{eqnarray}
S&=& \int_{-L}^0 dx \int_0^{T^{-1}} d\tau\,
\Big[{\partial_\tau  \phi_i \, \partial_\tau \phi_i \over 2}
+ {\partial_x \phi_i
  \, \partial_x \phi_i \over 2}
\nonumber \\
&&- \delta(x) \Gamma_\pm \cos(\sqrt{2 \pi} m^\pm_j \phi_j) \Big]
\label{tunaction}
\end{eqnarray}
The cosines result from a tunneling term of the form ${\psi_2}^\dagger \psi_1
+ {\rm h.c.}$, where the electron operator on an edge is a sum over operators
$O_{\bf m}$ with charge 1.  The form (\ref{tunaction}) assumes that each edge
has at most $n$ terms in the electron operator, so that possible phases in
the cosine term can be eliminated by translations $\phi \rightarrow \phi+a$.

At weak coupling the tunneling conductance scales with voltage $I \propto
V^\alpha$ with $\alpha = 3$, just as in tunneling between $\nu = 1$ and $\nu
= 1/3$, since the most relevant tunneling operators have scaling dimension 2,
but for strong coupling differences emerge, for example in the large-$V$
conductance, which is ${e^2\over 2 h}$ for $\nu = 1/3$ but ${4 e^2 \over 7
  h}$ for $\nu = 2/5$.  A simple picture of what happens at strong coupling
is that the coefficient of the cosine term $\Gamma_\pm$ becomes large,
trapping the $\phi$ fields at minima of the cosines; this corresponds to a
change from $N$ to $D$ boundary conditions on the neutral modes of the
combined system.

\begin{figure}
\centerline{\begin{picture}(250,105)
\multiput(20,20)(40,16){3}{\circle*{3}}
\multiput(20,52)(40,16){3}{\circle*{3}}
\multiput(20,84)(40,16){1}{\circle*{3}}
\multiput(100,20)(40,16){1}{\circle*{3}}
\put(10,36){1}
\put(20,22){\line(0,1){28}}
\put(22,20){\line(1,0){76}}
\put(55,10){$\sqrt{7}$}
\put(30,95){Original lattice}
\multiput(150,20)(12.8,32){3}{\circle*{3}}
\multiput(150,84)(12.8,32){1}{\circle*{3}}
\multiput(175.6,20)(12.8,32){3}{\circle*{3}}
\multiput(201.2,20)(12.8,32){3}{\circle*{3}}
\put(226.8,20){\circle*{3}}
\put(150,22){\line(0,1){60}}
\put(152,20){\line(1,0){21.6}}
\put(150,10){$2/\sqrt{7}$}
\put(140,50){2}
\put(187.12,55.2){\line(-2,5){10.4}}
\put(190.4,52){\line(1,0){21.6}}
\put(212.72,55.2){\line(-2,5){10.4}}
\put(177.6,84){\line(1,0){21.6}}
\put(190,65){$A$}
\put(150,95){Reciprocal lattice}
\end{picture}} 
\caption{
  The lattice of tunneling operators and its reciprocal lattice for tunneling
  between $\nu = 1$ and $\nu = 2/5$. The scaling dimension of an operator is
  its squared distance from the origin. The normalization for both lattices
  is such that length 1 corresponds to the self-dual radius ($\Delta = 1$).}
\label{figtwo}
\end{figure}
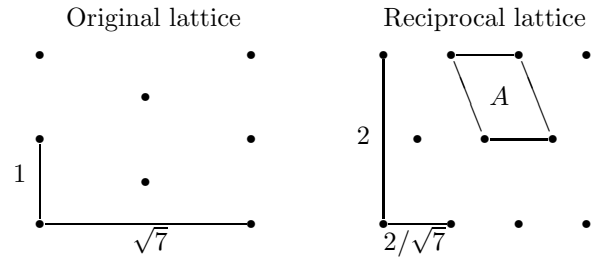

The joint minima in $(\phi_1,\phi_2)$-space of the two tunneling terms
$\cos(\sqrt{2\pi} m^\pm_i \phi_i)$ are given by a rectangular lattice
$\sqrt{2 \pi} [n_1 (2 / \sqrt{7},0) + n_2 (0,2)]$ plus a single basis vector
$\sqrt{2 \pi} (1/\sqrt{7},1)$.  Of course this lattice is also a Bravais
lattice with (nonorthogonal) vectors $(1/\sqrt{7},\pm 1)$.  At strong
coupling $\phi_1$ and $\phi_2$ are trapped at minima of the cosines
(Dirichlet boundary conditions), which are points on the reciprocal lattice
of the original operator lattice (Fig.~\ref{figtwo}).  The operator content
at strong coupling can be calculated via an instanton
expansion~\cite{schmid}: the operators correspond to tunneling paths between
different minima.  We will instead calculate the partition function to find
the (neutral) operator content, which verifies the instanton picture and
gives some additional information.  Because the instanton and
partition-function approaches agree, we expect the result to apply even if
strict conformal invariance is broken, e.g., by different mode
velocities.

We can write the conformal field theory (CFT) boundary
state~\cite{cardy,saleur} with
$\phi_1$ and $\phi_2$ pinned at minima of the cosines
as, up to translations of the rectangular lattice,
\begin{equation}
|B\rangle = C (|\phi_1 = \phi_2 = 0 \rangle
+ |\phi_1 = \sqrt{\frac{2 \pi}{7}}, \phi_2 = \sqrt{2 \pi}\rangle),
\label{boundstate}
\end{equation}
with $C$ some overall constant determined by the normalization of the
partition function.  Here the notation is that $|\phi_i = a\rangle$ is the
eigenstate of the operator $\phi_i(0,t)$ with eigenvalue $a$.
The partition function with both ends pinned can be directly calculated
from the boundary state (\ref{boundstate}) by a standard
technique~\cite{cardy} (a good pedagogical review of exactly this
type of calculation is in~\cite{saleur}):
\begin{eqnarray}
Z_{DD} &=& \langle B | \exp(-L H) | B \rangle \nonumber \\
&=& {1 \over \eta(q)^2} \Biggl[ \left( \sum_{n_1}q^{4 {n_1}^2 / 7} \right)
\left(\sum_{n_2} q^{4 {n_2}^2} \right) \nonumber \\
&&+\left(\sum_{n_1} q^{4 (n_1 + \frac{1}{2})^2 / 7} \right)
\left(\sum_{n_2} q^{4 (n_2 + \frac{1}{2})^2} \right) \Biggr].
\label{pfun}
\end{eqnarray}
Here $q = e^{-\pi/LT}$ is fixed by the system size and $\eta(q) =
q^{\frac{1}{24}}\prod_{n=1}^\infty (1 - q^n).$ There is an overall
constant from the charge mode (which always has free boundary
condition) which is ignored.  The partition function (\ref{pfun})
cannot be written as a product of the partition functions for two
bosons with well-defined radii, and the strong-weak coupling duality
does not reduce to independent duality transformations on $\phi_1$ and
$\phi_2$.  This results from the non-orthogonality of the basis
vectors in Fig.~\ref{figtwo}.

The partition function with both ends pinned allows us to read off the
scaling dimensions of neutral operators from the exponents of $q$.
The first scaling dimensions appearing are $\Delta = 4/7$, $\Delta =
8/7$, $\Delta = 4$, agreeing with lengths of vectors on the reciprocal
lattice.  The fact that the most relevant tunneling operator at strong
coupling has scaling dimension $\Delta = 4/7$ is additional evidence
that the limiting conductance is $\sigma_{hm} = 4/7$, since for
charge-unmixed FQH edges the most relevant tunneling operator has
$\Delta = \nu = \sigma$~\cite{wen}. The strongly coupled state is not
just a $\nu = 4/7$ quantum Hall state, however, because it has a
different spectrum of charge excitations away from the junction which
is not modified by the boundary interaction.

The above boundary state has a ``boundary entropy'' which has a natural
interpretation in terms of the lattice of minima.
The boundary entropy is defined
from the free energy in the $L \rightarrow \infty$ limit:
\begin{equation}
F = - T \log Z = -L f - T \log g_a - T \log g_b
\end{equation}
and $g_a$ and $g_b$ are the boundary contributions to the $T \rightarrow 0$
degeneracy.
The pinned boundary state $|B\rangle$ in (\ref{boundstate}) has
entropy
\begin{equation}
\Delta S_B = \log {g_B \over g_{\rm free}} = - \log A,
\quad A = l_1 l_2 / 2 = {2 \over \sqrt{7}}.
\label{bounddegen}
\end{equation}
Here $l_1,l_2$ are the lengths of the sides of the rectangular unit
cell and $A$ is the area of the true (nonrectangular) unit cell indicated in
Fig.~\ref{figtwo}.  Intuitively, a smaller unit cell means more minima
where $\phi_1, \phi_2$ can be trapped, and hence a greater boundary
entropy, in agreement with (\ref{bounddegen}).  Since $A < 1$ the
state $|B\rangle$ has greater entropy than the free state, so the
renormalization-group flow should be from $|B\rangle$ to the free
state, according to the principle that the RG reduces degrees of
freedom.  The same relationship $\Delta S_B= -\log A$ applies for
more complicated lattices (edges with more modes), where now $A$ is a
parallelogram in more dimensions.

Another way to calculate the conductance is from the operator content
at {\it weak coupling}, via a generalization of the
single-tunneling-operator result~\cite{fendley,chamon,moore2}.  For
tunneling between edges with one mode on each side, so that the
combined system has only one neutral mode ${\bf n}$, the conductance
difference between $N$ and $D$ boundary conditions is $\sigma_D -
\sigma_N = \frac{{e^*}^2}{h} \Delta^{-1}({\bf n}),$ where $e^*$ is the
charge transferred by the operator $O_{\bf n}$.  With several neutral
modes, this becomes a sum over an orthogonal basis of neutral modes:
\begin{equation}
\sigma_D - \sigma_N = \sum_i {{e^*_i}^2 \over h \Delta({\bf n}_i)}.
\label{condsum}
\end{equation}
The result $\sigma_D - \sigma_N = \sigma_{hm}$ follows from evaluating the
sum in the basis discussed above where only one neutral mode transfers charge.

For tunneling between $\nu=1$ and $\nu = 2/5$, it is instructive to look at
the conductance sum (\ref{condsum}) in a different basis. The $K$ matrix can
be rotated to be diag$(7/5,4,28)$, with only the first mode charged.  The
second mode is either of the two most relevant tunneling operators, which
have scaling dimension 2 and transfer a single electron just as in tunneling
between $\nu = 1$ and $\nu = 1/3$.  The third mode also transfers a single
charge and has its scaling dimension 14 fixed by orthogonality.  Then the
strong-coupling conductance is $\sigma = \frac{1}{2} + \frac{1}{14} =
\frac{4}{7}.$ In this basis the different strong-coupling conductances for
tunneling into $\nu = 1/3$ and $\nu = 2/5$ can be pictured as resulting from
the existence of an additional conduction channel in the $\nu = 2/5$ state.

The requirement that the conductance difference $\sigma = 4/7$ in
(\ref{condsum}) be the same when evaluated using the strong-coupling operator
content for tunneling between $\nu = 1$ and $\nu = 2/5$ suggests that the
operator of scaling dimension $4/7$ found above transfers charge $e^* =
\frac{4e}{7}$.  If so, it seems likely that the shot noise in the
``backscattering'' current for large $V$ ($I_{\rm B} = \frac{4}{7} V - I$) is
$S = 2 \frac{4e}{7} I_{\rm B}$.

Tunneling between edges of the same FQH state $\nu$ differs from the above in
that the physical strong-coupling fixed point does not correspond to $D$
boundary conditions on all neutral modes.  It was shown above that the
asymptotic large-voltage conductance is only sensitive to the boundary
condition on one neutral mode of the joint system, the ``charge transfer''
mode.  For point tunneling between similar edges, each with $n$ modes, $n$
neutral modes of the combined system (including the charge transfer mode)
acquire $D$ boundary conditions while the charge mode and $n-1$ neutral modes
stay with $N$.  This results from the ``folding'' symmetry in the action for
the case of $\nu$--$\nu$ tunneling~\cite{saleur}: we can form even and odd
combinations of the original fields, and only the even combinations couple to
the impurity.  The folding symmetry strictly exists only for exactly
identical edges and an idealized point impurity, so it may be possible to
access experimentally fixed points without this symmetry.

\begin{figure}

\epsfxsize=2.0truein
\vbox{\centerline{\epsffile{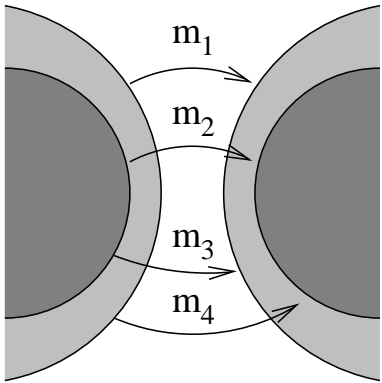}}
\caption{One representation of the four operators tunneling
quasiparticles between two $\nu = 2/5$ edges, each of which contains
two branches of edge modes.}}
\label{figthree}
\end{figure}

As an example of the above, consider tunneling between $\nu = 2/5$
states, where each subedge contains two modes (Fig. 3).  There is a
three-dimensional subspace of neutral operators of the combined system
(the four operators indicated in Fig. 3 are linearly dependent), and
$D$ boundary conditions on all neutral modes corresponds to lattice
duality in a three-dimensional subspace and leading operator
dimensions $(2/5,11/10,\ldots)$, while the fixed point preserving the
folding symmetry corresponds to duality in a two-dimensional subspace
(two neutral modes get $D$) and leading operator dimensions
$(2/5,3/5,\ldots)$, which are the dimensions of quasiparticle
tunneling operators.  The two modes which pass to $D$ boundary
conditions can be taken to be ${\bf m}_1$ and ${\bf m}_2$ in Fig. 3.
The possibility of different fixed points suggests that if the folding
symmetry is broken, e.g., by having multiple tunneling points, the
true strong-coupling point could correspond to the full
three-dimensional duality.

It is easy to prove that, with the restriction of folding symmetry,
the dual of electron tunneling between the same state $\nu$ is quasiparticle
tunneling.  The duality takes place in the $n$-dimensional subspace of
operators $({\bf m},-{\bf m})$ in our unfolded notation, since only
the $n$ even combinations couple to the impurity.
In the case of $\nu = 1$
to $\nu = 2/5$ tunneling discussed above, there is one neutral mode which can
be either $N$ or $D$ without altering the value of the conductance.  The
choice $D$ taken above corresponds to duality in a two-dimensional subspace,
while $N$ gives a leading tunneling operator of dimension $1/7$ and
transferred charge $2e/7$, which may be realizable if some tunneling operators
are tuned to zero.  We see that the
symmetries of a given tunneling problem help determine which of the
$\sigma = \sigma_{hm}$ fixed points is physically appropriate.

\section{Conclusions}
\label{sec:conclusions}

Our approach has been to study edge tunneling between general quantum
Hall states at three increasing levels of sophistication.  A simple
energy conservation argument gives an upper bound on the conductance
of a tunnel junction, and the Kubo formula applied to the $\chi$LL
theory of the edge modes indicates that the bound is saturated for
Dirichlet boundary conditions on the neutral mode which transfers
charge between the edges.  The same picture of strong-weak duality
previously obtained for the single-mode case also applies to the
general case, but with some new features: the duality in the
multiple-mode case does {\it not} reduce to independent dualities on
individual neutral modes of the combined system, but instead is a
multidimensional lattice duality.  For tunneling between specific
filling fractions, there can be several different fixed points with
the dissipationless value of conductance corresponding to dualities on
different sets of neutral modes; which fixed point is the physical
high-voltage limit can be determined from symmetry considerations.  In
sum, we have shown that the $\chi$LL model applied to hierarchical
edges has a consistent and physically reasonable strong-coupling limit
with interesting alterations from the exactly solvable single-mode
case.  The conductance, noise, and operator content at strong coupling
can be compared to possible experiments and to the predictions of
other theories.

The authors wish to thank E.~Fradkin and X.-G.~Wen for helpful comments.
Support was provided by the Fannie and John Hertz Foundation (J.~E.~M.),
NSF Grant DMR-98-76208 (P.~S. and C.~C.), and the Alfred P. Sloan
Foundation (C.~C.).

\end{document}